\documentclass [10pt,twoside]{article}

\usepackage{shrthnds}
\usepackage{cite}
\usepackage{graphicx}
\usepackage[usenames]{color}
\usepackage{subfigure}
\usepackage{setspace}

\usepackage{multirow}

\usepackage[hang,small]{caption}

\usepackage{amsmath}                  

\usepackage{geometry}
\usepackage{fancyhdr}
\usepackage{titlesec,titletoc}
  \titleformat{\section}{\Large\sf\bfseries}{\thesection}{1em}{}
  \titleformat{\subsection}{\large\sf\bfseries}{\thesubsection}{1em}{}

\title{\sf\bfseries \ntitle}
\author{ Pankaj Jain\footnote{email: pkjain@iitk.ac.in}, 
}
\date{}

\newcommand{\pghdr}{\footnotesize {P. Jain}  -- A flat space-time model
of the Universe}
\newcommand{\ntitle}{A flat space-time model of the Universe}

\begin{document}
\vspace{-3cm}
\maketitle
\vspace{-0.6cm}
\bc
\small{{}Department of Physics, IIT Kanpur, Kanpur 208 016, India
}
\ec

\bc\begin{minipage}{0.9\textwidth}\begin{spacing}{1}{\small {\bf Abstract:}
We propose a model of the Universe 
 based on Minkowski flat space-time metric. In this model the
space-time does not evolve. Instead the matter evolves such that all
the mass parameters increase with time. We construct a 
model based on unimodular gravity 
to show how this can be accomplished within the framework of
flat space-time. We show that the model predicts the Hubble law if the
masses increase with time. Furthermore we show that it fits the high $z$
supernova data in a manner almost identical to the standard Big Bang model. 
Furthermore we show that at early times the Universe is dominated by
radiative energy density. The phenomenon of recombination also arises
in our model and hence predicts the existence of CMBR. However a major
difference with the standard Big Bang is that the radiative temperature
and energy density does not evolve in our model. Furthermore we argue that
the basic motivation for inflation is absent in our model.

}\end{spacing}\end{minipage}\ec


\section{Introduction}
The observed Hubble's law has normally been interpreted in terms of the
expansion of the Universe. 
However within the framework of the Hoyle-Narlikar cosmological models,
it is also possible to
obtain this law within the framework of flat space-time 
\cite{Narlikar,Vishwakarma,Hoyle,Hoyle1,Narlikar1,Narlikar2}. 
The basic
idea is that all the particle masses evolve with time. Hence the transition
frequencies in the early Universe are different from the corresponding
frequencies observed in laboratory. If the masses increase with time, this
can explain the redshift of frequencies in the early Universe. 
Here we
show that this simple observation leads to a model of the Universe 
rather similar to the standard big bang model without invoking spatial
expansion. This important observation has also been made earlier 
\cite{Sasaki}. In this paper the authors argued that the standard 
Big Bang model can be reinterpreted in this manner, while all of it's
predictions remain unchanged. Here instead we point out that this may lead to 
 new avenues of research which might offer alternate solutions to
the problems faced by modern cosmology.  
The reinterpretation is so
dramatically different from the standard model, that it deserves a careful
investigation. In particular, within the alternate framework, the
Universe does not have a beginning, 
in contrast to the implications
of the standard Big Bang Model. The space-time exists for ever. 
Only the matter sector evolves. The origin of the Universe 
simply corresponds to a phase transition in the matter sector. 
We also show that the prediction for the
time elapsed in this model from the surface
of last scattering is very different from the prediction of the Big
Bang Model. Furthermore the concept of inflation may not have any relevance 
in this framework.

A model in which masses can effectively increase with time can be 
easily constructed with the framework of Minkowski space-time. 
The basic idea 
is rather simple. 
The construction is similar to the unimodular models discussed in 
Ref. \cite{Jain:2011,Jain:2012}. 
We assume
that there exists a background field, which we shall refer to as $\chi$.
In Ref. \cite{Jain:2011,Jain:2012} this field played the role
of the determinant of the metric. In the present case this is just another
scalar field. The coupling of this field to matter fields leads effectively
to time varying mass parameters. 

As mentioned above,
our proposal of time varying masses is similar to the proposal made
in the steady state models in order to explain the observed
redshift \cite{Narlikar}. However as we 
shall see, despite this similarity
our proposal is very different from those models. In fact, in
detail, our model turns out to be closer to the Big Bang model rather 
than the steady state model. This is despite the fact that 
the basic mechanism is very different.
In the Big Bang model the space expands.
Instead, in our model the length scales corresponding to matter, such
as size of atoms, size of nuclei, contract.  

\section{Mechanism}
We assume a model which leads to flat space-time and hence Lorentz metric.
Let us assume that all the dimensionful parameters scale with time. Hence
all the parameters with dimension of mass scale as,
\be
M(t) = f(t) M_0
\label{eq:Mt}
\ee
where $f(t)$ is a function of time. All the particle masses, the Planck
mass, the electroweak mass scale, the QCD mass scale follows this behaviour
with the same universal function $f(t)$.   
In fundamental units all other scales, such as energy, length, momentum
are related to the mass scale. Hence their evolution with time is also
specified by Eq. \ref{eq:Mt}. For example all length scales will 
decrease in time in proportion to $1/f(t)$. 
The function $f(t)$ is assumed to be a 
slow function of time, evolving on cosmological time scales. It plays
a role similar to the scale factor in standard Big Bang cosmology. 
Hence we can invoke adiabaticity and for most laboratory processes
this time dependence may be ignored.

We next explicitly construct a class of models, which lead to
Eq. \ref{eq:Mt} and hence cosmology without expansion, . 
We start by briefly reviewing the construction in \cite{Jain:2011,Jain:2012}
based on unimodular gravity \cite{Einstein,Anderson,Weinberg1,Ng1,Zee,Buchmuller}. 
We may split the metric as follows \cite{Jain:2011},
\begin{equation}
g_{\mu\nu}=\chi^2 \bar{g}_{\mu\nu}
\label{eq:gmunu}
\end{equation}
Let $\bar g$ denote the determinant of 
metric $\bar{g}_{\mu\nu}$.  
The unimodular constraint implies that $\bar g$ is equal to the 
determinant, $\eta$, of the Minkowski metric \cite{Jain:2011}. 
Hence the field $\chi$ plays the role of the determinant of the metric 
$g_{\mu\nu}$.
We shall only assume unimodular general coordinate invariance. In that
case the field $\chi$ is also a scalar. 

The action may be written in terms of the variables, 
$\chi$ and $\bar{g}_{\mu\nu}$, as follows \cite{Jain:2012},
\be
\mathcal{S} = \int d^4x \sqrt{-\eta}\left[{\chi^{2}\over \kappa} \bar R 
- {\xi\over \kappa} 
\bar g^{\mu\nu} \partial_\mu\chi\, \partial_\nu\chi\right] +\mathcal{S}_M 
\label{eq:action_gravity}
\ee
where $\mathcal{S}_M$ represents the matter action, 
$\kappa = 16\pi G$ and $G$ is the gravitational constant. 
If we demand full general coordinate
invariance, then $\xi=6$. 
If we only demand unimodular general coordinate invariance then
this parameter can take different values. 
Furthermore we can allow different functions of $\chi$ in the various 
terms in the action.
The matter action may be written as,
\be
\mathcal{S}_M = \int d^4x \sqrt{-\eta}\Bigg[
 {\chi^2\over 2} \bar g^{\mu\nu} \partial_\mu \phi \partial_\nu \phi - 
{\chi^{4 }\over 2} m^2\phi^2 - \Lambda\chi^{4} 
\Bigg]
\label{eq:matter}
\ee
Here we have only included a representative scalar field to represent the 
matter contribution. We may add fermion and vector fields, as 
discussed in \cite{Jain:2011,Jain:2012}. We have also included the contribution due to the
cosmological constant, $\Lambda$. 
The parameter $m$ represents the mass of the scalar field $\phi$  
Furthermore here also we can have different functions of $\chi$ in different
terms, as discussed in \cite{Jain:2011,Jain:2012}.
As discussed in \cite{Jain:2011,Jain:2012}, these models based on 
unimodular gravity yield predictions identical to Einstein's gravity
on the scale of solar system. This is because these deviate only in the
behaviour of the field $\chi$ which is relevant on cosmological scales.

The action, Eq. \ref{eq:action_gravity}, displays 
general covariance if $\xi=6$.  
It leads to the standard cosmic
evolution if we demand that invariant length scales are given by,
\be
ds^2 = \chi^2 \bar g_{\mu\nu} dx^\mu dx^\nu
\ee
The length scale invariant under general coordinate transformation is
$ds$, which has to be interpreted as the physical length. However if we
only demand unimodular general coordinate invariance, then the alternate
measure of length, ${d\bar s}$, defined as,
\be
{d\bar s}^2 =  \bar g_{\mu\nu} dx^\mu dx^\nu
\ee
is also a scalar. Hence this is also a well defined measure of distance. 
In our non-expanding Universe, we assume this to be the true measure of
distances. The resulting model then effectively leads to time varying
masses.  

We shall seek a solution to the equations of motion such that 
$\bar g_{\mu\nu}$ is equal to the Minkowski metric, $\eta_{\mu\nu}$.
Let us assume that $\chi$ is a slowly varying background field. 
Let us scale the scalar field field such that $\bar \phi=\chi\phi$. 
The resulting action in terms of this field can be written as, 
\be
\mathcal{S} = \int d^4x \sqrt{-\eta}\Bigg[{\chi^2\over \kappa}\bar R- { \xi\over\kappa }
 \bar g^{\mu\nu} \partial_\mu\chi\, \partial_\nu\chi
+ {1\over 2} \bar g^{\mu\nu} \partial_\mu \bar\phi \partial_\nu \bar\phi - 
{\chi^2\over 2} m^2\bar\phi^2 - \Lambda\chi^4 
 \Bigg]
\label{eq:action}
\ee
Here we have assumed adiabaticity and ignored terms of order $H/m$,
where $H$ is the Hubble constant. 
In the adiabatic limit, the matter field $\chi$ essentially leads to
particles with mass equal to $\bar m=\chi m$, which evolves with time.
We shall assume that masses of higher spin matter fields also scale in 
the same manner.

We point out that the kinetic energy term of the field $\chi$ is negative.
Phantom scalar fields
with negative kinetic energy terms have been introduced in many papers
in the literature, see, for example, 
\cite{Caldwell, Cline, Gibbons,Zhuravlev,Kujat}.

We can obtain the equation of motion for $\chi$
directly from Eq. \ref{eq:action}. 
We find \cite{Jain:2011,Jain:2012}
\be
2\chi\bar R
+ 2 \xi\bar g^{\mu\nu}\partial_\mu\partial_\nu\chi = \kappa T_\chi
\ee 
where $T_\chi$ is the contribution to this equation from all the
matter terms in $\mathcal S$, Eq. \ref{eq:action}, and the term proprotional
to $\chi^4$. Let us first assume that only radiation dominates. 
In the relativistic limit we find that,
\be
T_\chi = 0\ \ \ \ \ {\rm for\ radiation}
\ee
If the non-relativistic matter dominates the energy density of the
Universe, we then obtain, 
\be
T_\chi = \chi\left<m^2\bar\phi^2\right > \ \ \ \ \ {\rm for\ non-relativistic\ matter}
\label{eq:TchiNR}
\ee 
Here the expectation value is taken in a suitable thermal state, corresponding
to the state of matter in the Universe. As 
in the case of standard Big Bang, at leading order, we shall assume
that the matter is distributed homogeneously in the Universe. 
To be more precise, there exists a frame in which the distribution of matter,
on large distance scales, appears homogeneous and isotropic.
 Throughout this paper we assume
we are working in this frame.
In this case $\chi$ will only depend on time. 
Here we shall treat the field $\chi$ classically. However the matter
and radiation fields are quantized.
After
quantization the field $\bar\phi$ leads to particles which
pervade the Universe. Let us define the scaled mass, 
\be
\bar m = \chi m
\label{eq:barm}
\ee 
The action for $\bar\phi$ in this case becomes the standard free field
action with a time varying mass $\bar m$ \cite{Jain:2011}. We treat
this time variation adiabatically. Let us assume that at a given time the 
Universe has $N$ particles in volume V. Then the energy density at that
time
\be
\rho = \left<{\mathcal H}\right> = \left<\bar m^2\bar \phi^2\right> = {N\bar m\over V}
\ee 
Here $\mathcal H$ represents the Hamiltonian density.
Let us assume that the particles are not interacting. Hence their number
is fixed. The volume $V$ also does not change since space is not evolving.
Hence we find that 
\be
\rho = \left<\bar m^2\phi^2\right> \propto \bar m\propto \chi
\ee
This also implies that
\be
\left<m^2\bar\phi^2\right> \propto {\rho\over \chi^2} \propto {1\over \chi}
\ee
Hence we find that in the non-relativistic limit, $T_\chi$ is independent
of $\chi$. 
We may express the energy density in the non-relativistic limit as
\be
\rho_{NR} = \chi\rho_0
\ee
We shall set $\chi(t_0)=1$ 
 at the current time $t_0$. Hence $\rho_0$ represents the current
energy density.

For radiation dominated Universe, we obtain, 
\be
{d\chi\over dt} = \sqrt{\kappa\rho_0\over 6 }
\label{eq:evol_rad}
\ee
with $\rho=\rho_0$, independent of time.
For non-relativistic matter dominated Universe, we obtain,
\be
{d\chi\over dt} = \sqrt{\kappa\rho_0\chi\over 6}
\label{eq:evol_NR}
\ee
with $\rho=\rho_0\chi$.
Here we have set $\xi=6$.

Using Eq. \ref{eq:barm}, we identify the function $f(t)$ in 
Eq. \ref{eq:Mt} as, 
\be
f(t) = \chi(t)\ . 
\ee
Hence we find that we can obtain Eq. \ref{eq:Mt} purely within the
framework of Minkowski flat space-time metric. 
We point out that all matter fields, bosons or fermions, will be assumed to
have mass terms which are multiplied with the suitable power of $\chi$
such that the $\chi$ field in these terms can be absorbed by defining a
scaled mass $\bar m=\chi m$, as in the case of scalar field $\phi$. 
As we have explained above the Planck mass also scales in the same
manner. Hence, although all the masses are time dependent, the ratio
of any two scaled mass parameters remains fixed. 

The model with mass evolution leads to predictions identical to
physical observables such as luminosity distance, primoridal nuclear
abundances etc. as long as we use the action given
in Eq. \ref{eq:action_gravity} with $\xi=6$. 
However as we discuss in section 4, the nature of cosmic 
evolution is very different. An essential part of the standard big bang model
is the presence of initial singularity, when the Universe originated.
In the present model, however, the Universe, i.e. the space-time, 
was always present. The big bang singularity simply corresponds to a
phase transition in the matter sector. We discuss these points in detail
in section 4. 

Due to the evolution of masses, the frequencies of atomic transitions,
which are proportional to the mass of an electron, scale with time
as $f(t)$. Let $t_0$ denote the current time. 
Let $\nu_0$ be the frequency of an atomic transition
observed in laboratory. The frequency for the corresponding transition
 in the early Universe is denoted as $\nu(t)$, where $t$ is the time
when the light was emitted from a distant galaxy at redshift $z$. Hence
we obtain
\be
 {1\over 1+z} ={\nu(t)\over \nu_0} = {f(t)\over f_0} 
\label{eq:hubble}
\ee 
Here $f_0=f(t_0)=1$ and the first equality defines the redshift
$z$. By requiring that $f(t)$ increase with time, we find that atomic
transitions in the early Universe would be redshifted. Hence we obtain
the Hubble's law.  

\section{Luminosity Distance}
We next compute the luminosity distance in this model in order to fit
the high redshift supernova type 1a data. The model predicts an evolution
of these supernovae with time. It predicts that the 
luminosity of the supernova in the early Universe is smaller 
in comparison to the corresponding event ocurring today. This is because
the luminosity has dimensions of mass squared. This has to scale with
time since all dimensionful parameters in our model scale. Hence
we find that the luminosity, $L$, scales as,
\be
L(t) = f^2(t) L_0
\ee 
where $L_0$ is a constant equal to  
the luminosity of a current supernova explosion.  
Hence the flux received from such a source is 
\be
F = {L_0\over (1+z)^2 4\pi r^2} 
\ee
The luminosity distance $d_L$ is defined as,
\be
F= {L_0\over 4\pi d_L^2}
\ee
Hence we find
\be
d_L = (1+z)r
\label{eq:LumDis}
\ee
In the present flat space-time, the equation for $r$ is very simple. 
Using Eq. \ref{eq:hubble} we obtain
\be
r = t_0 - t = \int_{1/(1+z)}^1 {d\chi\over d\chi/dt}
\label{eq:time}
\ee
We can obtain $df/dt=d\chi/dt$ from the equation of motion of $\chi$. 
Substituting the resulting $r$ into Eq. \ref{eq:LumDis} gives the luminosity distance.
We now consider the equation of motion of $\chi$ for the case when only
non-relativistic matter and the vacuum energy term, proportional to $\chi^4$,
is present. We obtain 
\be
{d^2\chi\over dt^2} = {\kappa\over 2\xi} \left[\rho_0 + 4\Lambda\chi^3\right]
\label{eq:eomNRVAC}
\ee
Integrating this we obtain,
\be
{1\over 2}\left[{d\chi\over dt}\right]^2 = {\kappa\over 2\xi}
\left[\rho_0\ \chi + \Lambda\chi^4\right]
\ee
 Substituting this into $r$ we obtain the formula for luminosity
distance identical to that obtained in the standard Big Bang model. 
Hence we obtain a fit identical to the $\Lambda CDM$ model. 

One implication of the universal evolution function for all masses 
 is that the supernova light curve stretch factors will show the
standard scaling, as in the case of Big Bang cosmology \cite{Jain:2012}.
This is simply because of the fact that all dimensional parameters
scale with their appropriate mass dimension. The R-band supernova light
curve may be represented as \cite{Goldhaber},
\be
{I(t)\over I_{max}} = f_R\left((t-t_{max})/w\right) + b
\ee
where $I_{max}$ is the peak intensity and $t_{max}$ the corresponding time.
Here $w$ and $b$ are parameters. The parameter $w$ has dimensions of length.
Hence it will decrease with time in proportion to $1/f(t)$. Hence at
earlier time, corresponding to redshift $z$, we expect it to be larger
by a factor $(1+z)$, giving rise to the expected scaling $w=s(1+z)$,
where $s$ is independent of $z$.

\section{Cosmic Evolution}
In our model the space-time does not evolve. However the Universe does
evolve due to the time dependence of the background field $\chi$ which
effectively makes all the mass parameters evolve with time. 
Let us consider the time dependence of $\chi$ 
for the model with
$\xi=6$ in the non-relativistic limit. 
Let us first ignore the vacuum energy term. This term is relevant when
$\chi$ is very large.
In the absence of the vacuum energy term,
the solution is given by,
\be
\chi(t) = {\rho_0\over 4M^2} t^2 + C_1 t + C_2
\label{eq:chievol}
\ee
where $C_1$ and $C_2$ are integration constants.
During the radiation dominated phase, $\chi(t)$ has a linear dependence
on $t$, as given by Eq. \ref{eq:evol_rad}. 
Hence the integration constant $C_1$ gets related to the radiative energy
density. As we move back in time $\chi(t)$ continues to decrease. 
 At some stage we expect that $\chi(t)$ 
may become zero. This corresponds to the Big Bang singularity.
However in the present model this simply corresponds to a phase transition
in the matter sector. This may arise since the potential in our model
effectively varies with time. As $\chi\rightarrow 0$, the effective electroweak
scale becomes very small. Equivalently the Higgs mass term in the action
becomes very small and the background temperature contributions 
start to dominate. Hence this may lead to the electroweak phase transition.
Essentially at early time the electroweak symmetry may be restored 
leading to zero masses for all particles. 
 After electroweak phase transition, 
Higgs acquired a vacuum expectation value and all particles acquired
masses, which effectively evolve with time. At even earlier times
there might have been a GUT phase transition. 
The Big Bang singularity in our model, therefore, simply corresponds to
a phase transition in the matter sector.

 Hence we find that our model does not lead to any singularity
at early time, in contrast
to the singularity in the Big Bang model. 
Essentially there is no beginning since the background space-time 
always existed starting from time $t\rightarrow -\infty$. 
Hence the concept of age of the Universe has no
meaning in our model.

Let us now consider the solution when $\chi$ is very large. 
In this case the vacuum energy term dominates. This is relevant
at late times or at very early times, $t\rightarrow -\infty$. 
The solution at late time
can be expressed as,
\be
\chi = {T-t_0\over T-t}
\ee 
Here $T$ is an integration constant which is interpreted as some time larger 
than the current time. As $t\rightarrow T$, $\chi\rightarrow\infty$. Hence
in this limit all mass scales approach infinity and we reach a singularity.
This singularity arises due to presence of vacuum energy term. 
The model does not necessarily require the existence of a singularity, 
 in contrast to the Big Bang model. It is 
interesting to think of generalizations which may avoid the singularity,
arising due to the vacuum energy term.
One simple solution we propose here is to modify the $\chi^4$ term
in the action to $\chi^4\ g(\chi)$, where $g(\chi)$ is a function of
$\chi$ which is approximately equal to one for $\chi=\chi_0$ and 
rapidly approaches zero when $\chi\rightarrow \infty$. 
In this case the $\chi^4$ term will get suppressed as $\chi$ becomes
very large and the non-relativistic term will again dominate.
An alternate scenario is that $\chi\rightarrow 
\infty$ signals the onset of another phase transition.

Our model also differs from the Big Bang model
in terms of the 
behaviour of the time parameter. We should compare the time in our model
with the cosmic time in standard Big Bang model.
We can compute the age by using Eq. \ref{eq:time}. This should be interpreted
as the time elapsed since some early phase at large redshift.
We obtain,
\be
\Delta t = {3.43\over H_0}
\ee
which is more than three times the age obtained in the standard $\Lambda CDM$
model. Hence the time evolved turns out to be much larger despite the
fact that the luminosity distance is exactly the same.

We next point out that we do not require inflation in our model. 
Since the space does not expand, the concept of inflation does not
have any meaning in this model. The space 
in our model is flat due to the choice of Lorentz metric. The Universe 
does not expand and hence the horizon problem is also absent. 
We emphasize that the present model differs crucially from the Big Bang
model in this respect. The surface of last scattering does not correspond
to causally disconnected regions since the Universe existed for ever. 

\subsection{Cosmic Microwave Background Radiation}
A very important cosmological probe is the Cosmic Microwave Background Radiation (CMBR). We next determine whether it can arise naturally in our model also.
As we shall see it arises in our model in a manner very similar to that
in the standard Big Bang cosmology.

We first determine how the energy density of a background radiation 
field, which 
fills the Universe, will change with time in our model. The answer is, of
course, very simple. It does not evolve at all, as discussed in Section 2. 
If it is a black
body radiation corresponding to temperature $T$, the temperature remains
independent of time. Furthermore the energy density of such a radiation
field is independent of time. Hence we obtain,
\be
\rho_R = {\rm constant}
\label{eq:rhoR}
\ee
The evolution of the energy density of non-relativistic matter 
is given by,
\be
\rho_{NR} = f(t)\rho_0
\label{eq:rhoNR}
\ee 
where $\rho_0$ is the density today. 
 This shows that the energy density of 
non-relativistic matter increases with time. Hence in comparison
to relativistic matter, the energy density of non-relativistic matter
increases with time, exactly as in standard Big Bang cosmology. 
This implies that at some time in the past, the energy density of 
non-relativistic matter will fall below that of relativistic matter.
Hence radiative energy density dominates in the early Universe.

In early Universe all masses were very small. 
Hence all atomic and nuclear level splittings were smaller in comparison
to those observed today. However the frequency spectrum of background
radition is same as that observed today. This means that at sufficiently early 
times atoms could not exist. The radiation, even if it had peak 
frequency in microwaves, was too hot for atoms to exist. 
Hence at early times the density of neutral atoms was extremely small.

Let us now determine the condition for recombination, or formation of
atoms in the Universe. Here the calculation proceeds almost identically
to that in the Big Bang Model. We follow the treatment given in 
Ref. \cite{Dodelson}. We consider the reaction
\be
e^- + p \leftrightarrow H +\gamma
\ee  
Let $n_e$, $n_p$, $n_\gamma$ and $n_H$ denote the number densities of
electrons, protons, photons and neutral Hydrogen respectively. Here we
shall ignore the density of Helium. The Boltzmann
equation in the present case leads to \cite{Dodelson}
\be
{dn_e\over dt} = n^2_b\left<\sigma v\right>\left[{(1-X_e)\over n_b}
\left({\bar m_e T\over 
2\pi}\right)^{3/2} e^{-\epsilon_0/T} - X_e^2\right]
\label{eq:recom}
\ee
where $X_e = n_e/n_b$,
 $n_b=n_p+n_H$ is the baryon number density, 
$n_p\left<\sigma v\right>$ is the recombination rate, $T$ is
the temperature and $\epsilon_0= \bar m_e+ \bar m_p-\bar m_H$ is the binding energy of
Hydrogen. Here $\bar m_e$, $\bar m_p$, $\bar m_H$ denote the scaled 
mass parameters corresponding to electron, proton and Hydrogen atom
respectively.  We point
out that $n_e=n_p$ due to charge neutrality and $n_b$ is fixed due to
Baryon number conservation. 
Here we have assumed that all the free
neutrons have decayed by this time. 
Furthermore the temperature $T$ and the photon
number density $n_\gamma$ remains fixed in our model. 

As in the case of the Big Bang model we note that the left hand side is
of order $X_e H$, where $H$ is the Hubble parameter
\be
H = {df/dt\over f}
\label{eq:hubble1}
\ee
Here this represents the evolution of masses rather than space. We have 
assumed that $f(t)$ evolves very slowly with 
time and hence if the reaction rate $n_p\left<\sigma v\right>$ is
much larger than $H$, the term in brackets in Eq. \ref{eq:recom}
has to vanish. This gives us the condition for equilibrium \cite{Dodelson} 
\be
{X_e^2\over 1-X_e} = {n_\gamma\over n_b} {\sqrt\pi\over 2^{5/2}} \left({\bar m_e
\over T}\right)^{3/2} e^{-\epsilon_0/T}
\ee
Here we have used the standard formula for $n_\gamma$. The ratio
$n_\gamma/ n_b$ is extremely large, of the order of $10^{9}$. Let us
consider the time when $\epsilon_0\approx T$. In our model it is $\epsilon_0$
which is evolving and not $T$. At this time, $\bar m_e>>T$. Hence the right 
hand side is much larger than unity as long as $\epsilon_0\le T$. At 
much later time when $\epsilon_0>> T$ 
the exponential term becomes very small and $X_e\rightarrow 0$. Hence
we find that recombination occurs in our model in a manner almost 
identical to the Big Bang model. The precise time dependence of $X_e$ can
be deduced from Eq. \ref{eq:recom}.  
This will differ from the Big Bang cosmology due to difference in the
precise cosmic evolution.

\subsection{Different stages in the evolution of the Universe}
The Universe existed for all time with the scale parameter $f(t)=\chi(t)$
increasing with time. 
 At very early time all particles had zero mass. 
The Universe was filled with radiation at some
constant temperature. The potential function of fundamental physics 
effectively evolves due to the evolution of the field $\chi$. 
Due to the evolution of the potential phase transitions can occur.
Although the temperature of the radiative matter filling the Universe remains
constant over most times, it can change during a phase transition.
At some point the electroweak phase transition occurred due to this evolution. 
Beyond this time many particles acquire mass due to the Higgs vacuum
expectation value.  
All mass parameters increase with time due  
to the time dependence of $f(t)$. Analagously all length scales,
associated with matter, decrease
with time. 
 The temperature of the radiation is assumed to be the observed
CMBR temperature at current time, up to corrections which arise due to
annihilation of species which were earlier in thermal equilibrium. 
As the Universe evolves different species
fall out of equilibrium and either decay or evolve as primordial
relics. 
The condition for decoupling of a species from equilibrium
in our model is practically identical to that in the Big Bang model. Let
the species interact with the particles in the medium with the reaction 
rate equal to $\Gamma$.
As long as $H<\Gamma$, the species stays in equilibrium. Here 
$H$ is given by Eq. \ref{eq:hubble1}.   
When this condition is violated, the species decouples.
This is seen explicitly by use of the Boltzmann equation. Consider the
reaction $1+2\leftrightarrow 3+4$. Let $n_i$ denote the number density
of species $i$. Let \cite{Dodelson}
\be
n^{(0)}_i = e^{-\mu_i/T} n_i
\ee 
where $\mu_i$ is the chemical potential of species $i$ and $T$ is the 
temperature of the medium. 
The Boltzmann equation leads to \cite{Dodelson},
\be
{dn_1\over dt} = n_1^{(0)}n_2^{(0)}\left<\sigma v\right>\left[
{n_1 n_2 \over n_1^{(0)} n_2^{(0)}} 
-{n_3 n_4 \over n_3^{(0)} n_4^{(0)}} 
\right]
\label{eq:Boltz}
\ee
Let us assume that initially the species 1 is kept in equilibrium with the
medium due to it's reaction with species 2. During this time only the masses
of particles are undergoing evolution. This evolution takes place
on cosmic time scale. Hence the left hand side is of order $n_1H$. 
The right hand side is of order $n_1 \Gamma$, where $\Gamma$ is the reaction
rate $n_2\left<\sigma v\right>$. We assume that initially $\Gamma >> H$. 
Hence the equality in Eq. \ref{eq:Boltz} can be maintained only if 
\cite{Dodelson},
\be
{n_1 n_2 \over n_1^{(0)} n_2^{(0)}} 
={n_3 n_4 \over n_3^{(0)} n_4^{(0)}} 
\label{eq:equil}
\ee 
As the mass parameters in the theory increase, we expect that the cross 
section $\sigma$ will decrease. For example the weak cross sections are
proportional to the Fermi constant which has dimensions of inverse mass
squared. Hence it will decrease with time. Once the cross sections becomes
sufficiently small $H\approx \Gamma$ and Eq. \ref{eq:equil} no longer holds.
Beyond this time species 1 decouples from the medium.
As $H>>\Gamma$, species 1 stops interacting with the
medium and and hence evolves independently.

We have explicitly shown that at early times, the photons, electrons
and protons remain in thermal equilibrium. Hence the Universe
consists of plasma and photons are not be able to travel 
very far. At some stage, as the mass of particles grows, neutral atoms
form. The density of free electrons and protons  now falls 
with time. At some future time photons decouple as the density of
free charged particles becomes sufficiently small. These photons now
propagate freely in space and can be observed as the CMBR.

We point out that the phenomenon of primordial nucleosynthesis also 
proceeds in our model in a manner similar to the standard Big Bang. 
At early times we would have free protons and neutrons in equilibrium
with other particles. However as their masses and hence the binding energies 
increase, they start forming nuclei.
The primordial nucleosynthesis in our model also give
same results as in the standard Big Bang model for the model with
$\xi=6$. This
is because the main factor which controls the formation of nuclei is
$\exp(-B/T)$, where $B$ is the binding energy of nuclei and $T$ the
temperature of the medium. The dependence of the exponent on the scale
factor $f$ (or $\chi$) is exactly the same as in the standard Big Bang model.
The results may differ for the case when $\xi\ne 6$ or when the model
is further modified as discussed in \cite{Jain:2011,Jain:2012}.

\section{Discussion and Conclusions}
We have presented a model of the Universe based on flat space-time.
In this model the space-time is not dynamical. However the matter does
undergo evolution in this model. In particular we define effective masses
of all matter particles, which evolve with time. These lead to
the observed redshift of distant galaxies in a manner similar to that
proposed earlier within the framework of steady state theories
\cite{Narlikar}. However we show that the model also naturally 
gives the same dependence of luminosity distance on redshift as obtained
in the standard $\Lambda CDM$ model. In fact a contribution analogous
to the cosmological constant is required in our model also despite the 
fact that it has very different evolution in comparison to the 
$\Lambda CDM$ model. Our model, however, does not require inflation.
This is mainly because the Universe in our model always existed. There
was no beginning. There may have been a phase transition, such as the
electroweak phase transition, which lead to non-zero masses of all
the standard model particles. These masses have been undergoing cosmological
evolution ever since leading to phenomenon similar to what is 
expected in the hot big bang model. In particular 
primordial nucleosynthesis, recombination and decoupling happens in 
our model in a manner similar to the hot big bang model. Our 
model predicts the existence of CMBR in precisely the same manner
as the big bang model. The density perturbations in our model may 
be seeded by fluctuations in the background $\chi$ field.  
This will also lead to 
anisotropies in the CMBR. 
 We have  not 
considered this aspect of the model in this paper. 
This should be pursued in future research. 

We point out that although the fit to supernova data works out in precisely
the same manner as in big bang model, the time evolution in our model is
different. This leads to a significantly different estimate of the time elapsed 
between today and some early time corresponding to high redshift. 

The problems associated with the present day evolution of the Universe,
namely the fine tuning problem of the cosmological constant and the
coincidence problem of dark energy and dark matter is not solved in
the simple model we considered. We hope that the strikingly different
scenario we have presented may provide new avenues to solve these 
problems.

A fundamental question which is so far not answered is whether it is
possible to deduce from observations if the space is indeed expanding. 
This appears to be rather difficult, given that the two different 
interpretations give identical results for
the most significant
features of cosmological data, long interpreted as proof of expansion. 
However 
since our model requires only unimodular general covariance, it may be
interesting to explore models which are only invariant under this limited
transformations. These models may lead to observable predictions different
from the standard Big Bang model and hence may be tested.  
If such deviations from the standard Big Bang model can be ruled out then
it might imply that the determinant should be interpreted in the standard
manner. Furthermore 
inflation is not required by our model. Hence this might provide
another handle to test this proposal further.  

The idea that mass scales evolve with time appears to us to be much 
simpler in comparison to the notion of expansion of space. 
Furthermore it is satisfying that we can naturally obtain 
such an evolution in the
unimodular version of gravity, originally suggested by Einstein 
\cite{Einstein}.
We may point out that
Einstein also tried to modify the general theory of relativity by 
introducing the cosmological constant in order to
prevent the expansion of space \cite{Weinberg}. 
We have shown that this is possible, while agreeing with all cosmological
observations, if we
reinterpret the determinant of the metric.

\bigskip
\noindent
{\bf Acknowledgements} \\
I thank David Branch, Atul Jaiswal, Purnendu Karmakar, Gopal Kashyap, 
John Ralston and Mohammad Sami
for useful discussions.

\begin{spacing}{1}
\begin{small}

\end{small}
\end{spacing}
\end{document}